\documentclass{aa}
\usepackage{graphicx}

\newcommand{\gsim}{\,\raisebox{0.2em}{$>$}\!\!\!\!\!
\raisebox{-0.25em}{$\sim$}\,}
\newcommand{\lsim}{\,\raisebox{0.2em}{$<$}\!\!\!\!\!
\raisebox{-0.25em}{$\sim$}\,}

\begin{document}

   \title{Cosmic ray production in supernova remnants including
reacceleration: the secondary to primary ratio}

   \subtitle{}

   \author{E.G. Berezhko
          \inst{1}
          \and
          L.T. Ksenofontov
          \inst{1,2}
          \and
          V.S. Ptuskin
          \inst{3}
          \and
          V.N. Zirakashvili
          \inst{3}
          \and
         H.J. V\"olk
          \inst{4}}

   \offprints{H.J.V\"olk}

   \institute{Institute of Cosmophysical Research and Aeronomy,
                     31 Lenin Ave., 677891 Yakutsk, Russia\\
              \email{berezhko@ikfia.ysn.ru}
          \and
              Institute for Cosmic Ray Research, University of Tokyo,
                    Kashiwa, Chiba 277-8582, Japan\\
              \email{ksenofon@icrr.u-tokyo.ac.jp}
           \and
             Institute for Terrestrial Magnetism, Ionosphere
               and Radio Wave Propagation,
               Troitsk, Moscow region 142092, Russia\\
               \email{vptuskin@izmiran.rssi.ru}
               \email{zirak@izmiran.rssi.ru}
             \and 
               Max-Planck-Institut f\"ur Kernphysik,
                Postfach 103980, D-69029 Heidelberg, Germany\\
             \email{Heinrich.Voelk@mpi-hd.mpg.de}
             }

   \date{Received month day, year; accepted month day, year}
   
   \authorrunning{Berezhko et al.}
   \titlerunning{CR production in SNRs with reacceleration}

\abstract {We study the production of cosmic rays (CRs) in supernova
remnants (SNRs), including the reacceleration of background galactic
cosmic rays (GCRs) --- thus refining the early considerations by 
Blandford \& Ostriker (1980) and Wandel et al. (1987) --- and the effects 
of the
nuclear spallation inside the sources (the SNRs). This combines for the
first time nuclear spallation inside CR sources and in the diffuse
interstellar medium, as well as reacceleration, with the injection and
subsequent acceleration of suprathermal particles from the postshock
thermal pool. Selfconsistent CR spectra are calculated on the basis of the
nonlinear kinetic model. It is shown that GCR reacceleration and CR
spallation produce a measurable effect at high energies, especially in the
secondary to primary (s/p)  ratio, making its energy-dependence
substantially flatter than predicted by the standard model.
Quantitatively, the effect depends strongly upon the density of the
surrounding circumstellar matter. GCR reacceleration dominates secondary
CR production for a low circumstellar density. It increases the expected
s/p ratio substantially and flattens its spectrum to an almost
energy-independent form for energies larger than 100~GeV/n if the
supernovae explode on average into a hot dilute medium with hydrogen
number density $N_\mathrm{H}=0.003$~cm$^{-3}$. The contribution of CR
spallation inside SNRs to the s/p ratio increases with increasing
circumstellar density and becomes dominant for $N_\mathrm{H}\gsim
1$~cm$^{-3}$, leading at high energies to a flat s/p ratio which is only
by a factor of three lower than in the case of the hot medium.
Measurements of the boron to carbon ratio at energies above 100~GeV/n
could be used in comparison with the values predicted here as a
consistency test for the supernova origin of the GCRs.
    }
   \maketitle
%

\section{Introduction}

It is a widely accepted hypothesis that supernova remnants (SNRs) are the
main sources of cosmic rays (CRs) in the Galaxy.  Supernovae occur
randomly in the Galaxy (although with some correlation in space) and have
a complicated time evolution. The selfconsistent nonlinear kinetic theory
of diffusive shock acceleration nevertheless gives rather definite
predictions on the shape, the amplitude and the time evolution of the
spectra of CRs accelerated at the shock during the different phases of SNR
evolution (Berezhko et al. \cite{byk96}; Berezhko and V\"olk
\cite{bv97};  Berezhko and Ksenofontov \cite{bk99}).

Released from SNRs, CRs experience energy-dependent diffusion and,
possibly, large scale convective transport in the interstellar medium
(ISM). During their propagation in the Galaxy, the primary relativistic
nuclei interact with interstellar gas nuclei and produce lighter secondary
relativistic nuclei as a result of nuclear spallation. The secondary
nuclei have approximately the same energy per nucleon as the parent
primaries.

Boron nuclei represent an example of pure secondaries and the
Boron-to-Carbon (B/C) ratio is an example of the secondary-to-primary
(s/p) ratio in CRs. The abundance of secondary nuclei, measured over a
wide energy range, yields important information on the process of CR
transport in the Galaxy (e.g. Berezinskii et al. \cite{brz90}).  Up to
approximately $100$~ GeV/n the measured power law spectrum of secondary
nuclei is steeper than the spectrum of primary nuclei. This leads to the
conclusion that primary GCRs are mainly produced in compact sources (SNRs
for example) and that their spectrum $n(\epsilon_\mathrm{k})\propto
\epsilon_\mathrm{k}^{-\gamma_\mathrm{g}}$, for kinetic energies 
$\epsilon_\mathrm{k}\gg1$~GeV/n, is
steeper than the spectrum, $N(\epsilon_\mathrm{k})\propto 
\epsilon_\mathrm{k}^{-\gamma_\mathrm{s}}$
of the particles produced in the sources, the so-called source cosmic rays
(SCRs): the GCR power law index $\gamma_\mathrm{g}=\gamma_\mathrm{s} +\mu$ 
is considerably
larger than the SCR power law index $\gamma_\mathrm{s}$ with a value $\mu$ r
anging between $\mu=0.3$ and 0.7.
 
The diffusion of relativistic particles in the Galaxy, after they have
left the compact sources (assumed here to be SNRs), may be accompanied by
an additional distributed reacceleration in the turbulent ISM. This
reacceleration can not be strong for particles with energies $1$ GeV/n
$\lsim \epsilon_\mathrm{k}\lsim100$ GeV/n as evident from the observed
energy-dependent decrease of the abundance of secondaries in this energy
range (Hayakawa \cite{haya69}).  Some weak reacceleration is possible and may in
fact explain the observed energy dependence of the s/p ratios with a
characteristic peak at about $1$ GeV/n (see Jones et al. \cite{jlpw01}, and
references therein). In the ''minimal'' model, the stochastic
reacceleration is produced by the same randomly moving inhomogeneities
which are assumed to be responsible for the spatial diffusion of CRs in
the galactic magnetic field. The velocity of these inhomogeneities is
close to the Alfv\'en velocity $c_\mathrm{a}\sim30$ km/s. The characteristic time
of distributed reacceleration is approximately equal to the escape time
from the Galaxy for particles with magnetic rigidities of about 1~ GV (the
magnetic rigidity is $R=pc/Ze$ , where $p$ is the particle momentum, $Ze$
denotes the particle charge, and $c$ is the speed of light). Escape
dominates over reacceleration at high rigidities $R>>1$ GV, so that the
effect of distributed reacceleration in the interstellar turbulence is
weak at high energies. In detail the importance of distributed
reacceleration for the formation of the GCR spectrum remains unclear. It
may well be that the peaks in the s/p ratios are caused by a specific
dependence of the CR diffusion on energy or by the effect of large scale
convection on the transport of low energy CRs (Jones et al. \cite{jlpw01}).  It
should be stressed however that the distributed reacceleration through
interstellar Alfv\'enic turbulence is negligible at energies
$\epsilon_\mathrm{k}\gsim20$~GeV/n in all versions of the galactic CR
propagation.
 
The scenario described is valid for the major part of the GCRs which left
the source regions and do no more interact with SNR shocks propagating in
the ISM.  However, a wandering energetic particle has a finite probability
to again meet a SNR shock which efficiently accelerates CRs, and therefore
to undergo strong reacceleration. The background primary and secondary CR
nuclei experience strong reacceleration and significant transformation of
their energy spectra after interaction with high velocity shocks. Strongly
reaccelerated primaries and secondaries at the shock have flat spectra
typical of the source spectrum. These particles are mixed with and diluted
by a large amount of background GCRs after their release from their parent
SNR. In principle they may become recognizable above some energy as a flat
component of galactic secondary nuclei.
 
The effect described above was studied by Wandel et al. (\cite{wel87}) 
(see also
the original paper by Blandford \& Ostriker \cite{bo80}) who found that a
flattening of the s/p ratio can be expected at high energies due to
reacceleration by strong SNR shocks. At the same time, applied to an
actual galactic ISM, this model leads to the conclusion (Wandel
\cite{wan88}) that
one has to expect a reacceleration effect mainly due to old and weak SNR
shocks of Mach numbers $M<2$ which also essentially influence the CR
spectra at GeV energies and make the CR mean escape length from the Galaxy
as small as 3 to 5~g cm$^{-2}$. These conclusions were based
on a simplified plane wave approach. All the SN shocks were considered to
be of the same strength and the final predictions depend substantially on
this suggested strength.

We note also that it is hard to believe that such weak shocks are even
able to accelerate relativistic particles in the same manner as a plane
shock. Indeed, to be efficiently accelerated by a spherical shock of size
$R_\mathrm{s}$ and speed $V_\mathrm{s}$, particles should have a sufficiently 
small diffusion coefficient (e.g. Berezhko \cite{ber96})
\begin{equation}
\kappa \lsim 0.1 R_\mathrm{s}V_\mathrm{s}.  \label{eq1}
\end{equation}
Shocks of size $R_\mathrm{s}\approx 30$~pc and speed $V_\mathrm{s}\approx 
300$~km/s (Wandel
\cite{wan88}) give the restriction $\kappa \lsim 3\times 10^{26}$~cm$^2$/s. Such a
small value can only be produced near the shock front by selfexcited
Alfv\'en waves, because the CR diffusion coefficient due to interstellar
turbulence exceeds this value even at GeV energies (e.g. Berezinskii et
al. \cite{brz90}). On the other hand the spectrum of particles accelerated by
such weak shocks is so steep, $n\propto p^{-\gamma}$ with $\gamma \approx
4$, that, even if extended into the relativistic energy range $p>mc$, it
contains too small an amount of energy to substantially amplify the ISM
turbulence. This means that efficient acceleration of relativistic
particles in SNRs stops earlier, at higher Mach numbers $M\approx 4$, when
the shock compression ratio drops significantly below the asymptotic
nonrelativistic value $\sigma =4$, and freshly accelerated particles
become unable to provide a high level of turbulence near the shock front.
Therefore it is assumed that at this age $t=T_\mathrm{SN}$, SNRs release all
previously accelerated CRs into the surrounding ISM.

The energy increase of relativistic CRs upon encountering the SNR of age
$t>T_\mathrm{SN}$ is negligible because their relatively large diffusive length
$l=\kappa/V_\mathrm{s}>0.1 R_\mathrm{s}$ reduces their energy gain at the shock front
almost to the adiabatic energy loss in the expanding downstream region.
Therefore we neglect the acceleration and/or reacceleration of CRs by
these old SNRs.

Since according to the above physical arguments
the maximum age and the corresponding number of SNRs which
effectively accelerate CRs is presumably
much smaller than assumed by Wandel (\cite{wan88}),
the effect of reacceleration of GCRs is also much smaller. Nevertheless it
is still important, especially at high energies for the case of a low-density
ISM.

In this paper we shall in addition take the effect of nuclear spallation inside
the sources into account. The energy spectrum of these source secondaries is
harder than that of reaccelerated secondaries. Therefore it plays dominant role
at high energies for a high-density ISM.

Note that the CR source spectrum $N(\epsilon_\mathrm{k})$ which is the
resultant overall spectrum of CRs accelerated during the whole active
period of SNR evolution depends on the GCR spectrum
$n(\epsilon_\mathrm{k})$ since not only the injection of suprathermal
particles leads to the formation of $N(\epsilon_\mathrm{k})$ but the
reacceleration of existing background GCRs contributes as well. These
spectra $N(\epsilon_\mathrm{k})$ should therefore be determined through a
selfconsistent solution of the equations which describe the GCR
propagation in the Galactic confinement volume with all accompanying
effects, together with the equations which describe the SNR evolution and
the CR acceleration. Fortunately this rather complicated problem can be
simplified without loss of accuracy. Indeed, due to the steep GCR spectrum
at relativistic energies $\epsilon_\mathrm{k}\gg 1$~GeV/n only the
injection of GCRs with energies $\epsilon_\mathrm{k} \lsim 1$~GeV/n plays
a role for the formation of $N(\epsilon_\mathrm{k})$. Since the GCR
spectra $n(\epsilon_\mathrm{k})$ at these subrelativistic energies are
known from experiment, we use them as an physical input quantities, like
other ISM parameters, in our selfconsistent nonlinear kinetic model for CR
acceleration in SNR, so that $N(\epsilon_\mathrm{k})$ can be considered as
consistently determined together with the GCR spectra
$n(\epsilon_\mathrm{k})$.

We also want to emphasize that the shape of the overall CR spectrum
$N(\epsilon_k,t)$ differs at each stage of SNR evolution significantly
from the spectrum of freshly accelerated CRs $n(\epsilon_k,t)$, especially
during the late Sedov phase. In this late phase, when the shock is
relatively weak and thus CR backreaction not significant, the freshly
produced CR spectrum has a pure power law form $n\propto p^{-\gamma}$,
bounded by an exponential cutoff at some maximum momentum $p_m(t)$. At the
same time, the overall CR spectrum is influenced by the nonlinear shock
modification during the previous active period of the SNR evolution, by
the adiabatic cooling of CRs in the downstream region, and by the
so-called escape effect (see Berezhko et al. \cite{byk96} for details) and
therefore has a much more complicated form. Only at very low energies
$\epsilon_k \lsim 1$~GeV it has a form similar to $n(p,t)$, whereas at
higher energies it becomes progressively flatter and is bounded by the
cutoff momentum $p_{max}=p_m(t_0)>p_m(t)$, a value which is determined by
the very early Sedov phase (Berezhko \cite{ber96}). This is an additional
and significant difference between our model, based on the kinetic
nonlinear time-dependent approach, and Wandel's (\cite{wan88}) estimates,
based on a plane-wave approximation, which does not include these
important physical effects.

In section 2 we present the relevant relations within the standard leaky
box model. In section 3 the GCR injection problem within the nonlinear
kinetic model for CR production in SNRs is described and the relevance of
secondary nuclei production in SNRs for the resultant source spectra
$N(\epsilon_\mathrm{k})$ and for the s/p ratio are estimated. Numerical results for
the boron and carbon spectra produced in SNR and the expected B/C ratio
are presented and discussed in section 4.  The final section contains the
main conclusions.

\section{Source contribution to the secondary to primary ratio}

We shall use the simple leaky box model to describe the transport and
nuclear fragmentation of comic rays in the Galaxy. It is well known that
the more adequate diffusion model of cosmic ray propagation with an
extended flat halo gives essentially the same results as the leaky box
model for the abundances of the stable primary and secondary nuclei. 
The approximate equivalence of these two models
holds provided that cosmic ray nuclear fragmentation and reacceleration
occur in regions much thinner than the cosmic ray halo size. This last
condition is fulfilled in the Galaxy where the disk thickness of the order
$0.5$ kpc is much smaller than the halo thickness (which is at least an
order of magnitude larger) (Berezinskii et al. \cite{brz90}).

For a simple estimate we consider the case when the secondary nuclei of a
species "s" are the result of spallation of some primary CR nuclei of
species "p", and the contribution of all other primary parent nuclei can
be neglected. Within this approach the differential number density of some
primary CR species $n_\mathrm{p}(\epsilon_\mathrm{k})$ is determined by the 
balance equation
\begin{equation}
\frac{n_\mathrm{p}}{\tau_\mathrm{c}}=\frac{N_\mathrm{p}}{V_\mathrm{c}}
\nu_\mathrm{SN}-\sigma_\mathrm{p} N_\mathrm{g} v n_\mathrm{p}, 
\label{eq2}
\end{equation}
where $V_\mathrm{c}$ is the residence volume occupied by the GCRs,
$\tau_\mathrm{c}$ is the
mean residence time in this volume $V_\mathrm{c}$, $\sigma_\mathrm{p}$ 
is the spallation
cross section, $N_\mathrm{p}(\epsilon_\mathrm{k})$ is the total differential 
number of CRs
created during the entire evolution of a single SNR (overall CR spectrum),
$\nu_\mathrm{SN}$ is the galactic SN explosion rate, $v$ is the particle speed,
$N_\mathrm{g}=\rho_0/m_\mathrm{p}$ is the ISM number density, $\rho_0$ is 
the ISM density,
and $m_\mathrm{p}$ denotes the proton mass.

Within the same approach the number density of secondary nuclei
$n_\mathrm{s}$
obeys the equation
\begin{equation}
\frac{n_\mathrm{s}}{\tau_\mathrm{c}}=\frac{N_\mathrm{s}}{V_\mathrm{c}}
\nu_\mathrm{SN}+\sigma_\mathrm{ps}N_\mathrm{g} v
n_\mathrm{p}-\sigma_\mathrm{s} N_\mathrm{g} v n_\mathrm{s},
\label{eq3}  
\end{equation}
where $\sigma_\mathrm{ps}$ is the partial cross section for creation of secondary
nuclei in hadronic collisions of primaries with the ISM gas nuclei, $N_\mathrm{s}$
is the total number of secondary CRs created during the evolution of a
single SNR.  Compared with the standard leaky box model the above equation
contains an additional source term which describes the production of
secondaries in SNRs. In the above equations all variables are functions of
the same kinetic energy per nucleus $\epsilon_\mathrm{k}$.

One can easily find the s/p ratio from the above two equations:
\begin{equation}
\frac{n_\mathrm{s}}{n_\mathrm{p}}=
\frac{\sigma_\mathrm{ps}N_\mathrm{g}v\tau_\mathrm{c}}{1+\sigma_\mathrm{s} 
N_\mathrm{g} v\tau_\mathrm{c}}+
\frac{N_\mathrm{s}}{N_\mathrm{p}}
\frac{(1+\sigma_\mathrm{p} N_\mathrm{g} v\tau_\mathrm{c})}{(1+\sigma_\mathrm{s} 
N_\mathrm{g} v \tau_\mathrm{c})}.
\label{eq4}
\end{equation}
Since the residence time $\tau_\mathrm{c}$ in the GCR confinement volume is a
decreasing function of CR energy for $\epsilon_\mathrm{k}\gg 1$~GeV/n, we have
\begin{equation}
\sigma N_\mathrm{g} v \tau_\mathrm{c}\ll 1,
\label{eq5}
\end{equation}
and the expression for the ratio can be drastically simplified:
\begin{equation}
\frac{n_\mathrm{s}}{n_\mathrm{p}}=
\frac{\sigma_\mathrm{ps} x}{m_\mathrm{p}}+
\frac{N_\mathrm{s}}{N_\mathrm{p}},
\label{eq6}
\end{equation}
where we have introduced the escape length 
\begin{equation}
x=m_\mathrm{p}N_\mathrm{g}v\tau_\mathrm{c},
\label{eq7}
\end{equation}
in the form of the mean matter thickness traversed by GCRs in the course
of their random walk in the Galaxy. Expression (\ref{eq6}) shows that apart from
the usual term proportional to the escape length $x$, the s/p ratio
contains an additional term which describes the source contribution in the
production of secondaries. Since the primary spectra 
$N_\mathrm{p}(\epsilon_\mathrm{k})$,
and the secondary spectra $N_\mathrm{s}(\epsilon_\mathrm{k})$, 
presumably produced in SNRs,
have a similar energy dependence, we expect the term 
$N_\mathrm{s}/N_\mathrm{p}$ to dominate
at sufficiently high energies in the s/p ratio.

One can also get an expression for the escape length 
$x$ from Eq. (\ref{eq4}) in
terms of the observed s/p ratio $r_\mathrm{g}=n_\mathrm{s}/n_\mathrm{p}$ 
and the source s/p ratio
$r_\mathrm{s}=N_\mathrm{s}/N_\mathrm{p}$:
\begin{equation}
x=\frac{m_\mathrm{p}(r_\mathrm{g}-r_\mathrm{s})}
{\sigma_\mathrm{ps}+r_\mathrm{s}\sigma_\mathrm{p}- r_\mathrm{g}\sigma_\mathrm{s}}.
\label{eq8}
\end{equation}

In the simplest case when the source produces only primary CRs $(r_\mathrm{s}=0)$,
we have the usual expression
\begin{equation}
x=m_\mathrm{p}/(\sigma_\mathrm{ps} /r_\mathrm{g} -\sigma_\mathrm{s}),
\label{eq9}
\end{equation}
which makes it possible to extract the value of $x$ from the measured s/p
ratio $r_\mathrm{g}$. Note that in the general case, when an unknown number of
secondary CRs are produced in the CR sources, there is no one to one
correspondence between $r_\mathrm{g}$ and $x$.

\section{Acceleration and reacceleration of CRs in SNRs}
There are two different suprathermal particle populations in the ISM which
are injected into the diffusive shock acceleration process in SNRs. The
first and most general one is the injection of some fraction of the
postshock thermal particle distribution.  It occurs for all ions present
in the background medium and usually supplies enough particles to convert
an significant part of the SN shock energy into that of an energetic
particle population.

The second possibility is the acceleration of pre-existing GCR particles
which have a sufficiently high energy $\epsilon_\mathrm{k} \gsim 100$~MeV/n 
so that
they participate naturally into the acceleration process. Note that for
those elements which are strongly underabundant in the thermal ISM, like
Li, Be, B, this is the only practical possibility for acceleration. To
distinguish these two different injection mechanism we use here the term
``acceleration'' for the first case and ``reacceleration'' for the second.

We shall employ here a simple CR injection model, in which a small
fraction $\eta$ of the incoming thermal protons is instantly injected at
the gas subshock with a speed that exceeds the postshock gas sound speed
$c_\mathrm{s2}$ by a factor $\lambda>1$ (Berezhko et al. \cite{byk96}):
\begin{equation}
N_\mathrm{inj}= \eta N_\mathrm{g1} , ~ 
p_\mathrm{inj}=\lambda m c_\mathrm{s2}.  
\label{eq10}
\end{equation} 
Here $N_\mathrm{g}=\rho/m_\mathrm{p}$ is the gas number density, 
and the subscripts 1(2) refer to the point just ahead (behind) the shock.

GCR nuclei (primary and secondary) have a spectrum which in the
subrelativistic region rises as a function of $\epsilon_\mathrm{k}$ due
to ionization and nuclear losses. The spectrum has a maximum at
$\epsilon_\mathrm{k} =\epsilon_\mathrm{GCR}\approx 600$~MeV/n and then
falls of in a steep power law for higher energies.  Therefore in the
case of reacceleration it is assumed that the existing GCR population
is injected at the SN shock front into the diffusive acceleration with
this energy
\begin{equation}
p_\mathrm{inj}= p_\mathrm{GCR},~
N_\mathrm{inj}=N_\mathrm{GCR}.
\label{eq11}
\end{equation}
Here $N_\mathrm{GCR}$ is the total number of GCR nuclei per unit volume and 
$p_\mathrm{GCR}$
is their mean momentum, that corresponds to $\epsilon_\mathrm{GCR}$.

Since the primary CRs come from both injection mechanisms, we can estimate
their relative role for the final CR production in SNRs. In order to do
that we compare the overall CR spectra 
(i.e. the particle numbers $dN$ in the momentum interval $dp$)
$N(p)$ and $N^\mathrm{re}(p)$ which
correspond to these two injection mechanisms. The overall spectrum
of CRs reaccelerated in each single SNR at $p\ge p_\mathrm{GCR}$ is determined by
the simple relation
\begin{equation}
N^\mathrm{re}=\frac{V_\mathrm{SN}N_\mathrm{GCR}} { p_\mathrm{GCR}}
\left( \frac{p}{p_\mathrm{GCR}}
\right)^{-\gamma_\mathrm{s}},
\label{eq12}
\end{equation}
where $V_\mathrm{SN}$ is the volume of the SNR at the latest evolutionary stage
where efficient CR acceleration can occur. For the sake of simplicity we
assume that the overall spectrum of CRs produced during the active period
of SNR evolution has a pure power law form with index $\gamma_\mathrm{s}$.

After their release from the parent SNRs these reaccelerated CRs occupy
the confinement volume $V_\mathrm{c}$ more or less homogeneously with a number
density $n_\mathrm{GCR}^\mathrm{re}$ that can be found from the leaky box equation
(\ref{eq2})(we neglect here the collision term):
\begin{equation}
n_\mathrm{GCR}^\mathrm{re}/\tau_\mathrm{c}=N^\mathrm{re}\nu_\mathrm{SN}/V_\mathrm{c}.
\label{eq13}
\end{equation}
Since GCRs have a power law spectrum $n_\mathrm{GCR}\propto p^{-\gamma_\mathrm{g}}$,
their total number density is 
\begin{equation}
N_\mathrm{GCR}=n_\mathrm{GCR}(p_\mathrm{GCR})p_\mathrm{GCR}/(\gamma_\mathrm{g}-1)
\label{eq14}
\end{equation}
and therefore we can write
\begin{equation}
\frac{n_\mathrm{GCR}^\mathrm{re}}{n_\mathrm{GCR}}=
\frac{\nu_\mathrm{SN}\tau_\mathrm{c} V_\mathrm{SN}}
{(\gamma_\mathrm{g}-1)V_\mathrm{c}}
\left( \frac{p}{p_\mathrm{GCR}}
\right)^{\gamma_\mathrm{g}-\gamma_\mathrm{s}}.
\label{eq15}
\end{equation}

Note that this relation is valid for primary as well as for secondary
elements. The only difference between them is the different momentum
dependence of the ratio $n_\mathrm{GCR}^\mathrm{re}/n_\mathrm{GCR}$. 
For primary elements as it
follows from the equation (\ref{eq2})
\begin{equation}
\gamma_\mathrm{g}=\gamma_\mathrm{s} +\mu,
\label{eq16}
\end{equation}
where power law index $\mu$ determines the momentum dependence of the residence
time $\tau_\mathrm{c} \propto p^{-\mu}$. Therefore the ratio 
$n_\mathrm{GCR}^\mathrm{re}/n_\mathrm{GCR}$ is
independent of $p$ in this case.

The spectrum of secondary elements, produced in the Galactic disk 
according to the equation (\ref{eq2}),
is considerably steeper, because
\begin{equation}
\gamma_\mathrm{g}=\gamma_\mathrm{s} +2\mu,
\label{eq17}
\end{equation}
and therefore in this case the contribution of GCR reacceleration
\begin{equation}
n_\mathrm{GCR}^\mathrm{re}/n_\mathrm{GCR} \propto (p/p_\mathrm{GCR})^{\mu}
\label{eq18}
\end{equation}
increases with energy.

The overall CR spectrum is formed during the active period of SNR
evolution which lasts up to the time when the SN shock becomes too weak
to accelerate efficiently a new portion of freshly injected particles.
According to calculations presented below this final stage corresponds
to a shock size $R_\mathrm{s}\approx 10R_0$ in the case of an ISM with
hydrogen number density $N_\mathrm{H}=1$~cm$^{-3}$, and
$R_\mathrm{s}\approx3R_0$ in the case of a hot ISM with
$N_\mathrm{H}=0.003$~cm$^{-3}$, where $R_0=(3M_\mathrm{ej}/4\pi
\rho_0)^{1/3}$ is so the called sweep-up radius, $M_\mathrm{ej}$ is the
ejecta mass, and $\rho_0=N_\mathrm{g}m_\mathrm{p}= 1.4
N_\mathrm{H}m_\mathrm{p}$ is the ISM density. Taking into account that
$\tau_\mathrm{c}/V_\mathrm{c}=\tau_\mathrm{g}/V_\mathrm{g}$, where
$\tau_\mathrm{g}$ is the residence time in the galactic disk volume
$V_\mathrm{g}=2.5\times 10^{66}$~cm$^3$, and
$\nu_\mathrm{SN}=1/30$~yr$^{-1}$, we have for energies about 5~GeV,
where $\tau_\mathrm{g}=4.6\times 10^6$~yr, a value
$n_\mathrm{GCR}^\mathrm{re}/n_\mathrm{GCR}=0.04$ for
$N_\mathrm{H}=1$~cm$^{-3}$, and $n_\mathrm{GCR}^\mathrm{re}
/n_\mathrm{GCR}=0.3$ for $N_\mathrm{H}=0.003$~cm$^{-3}$. One can see
that in an ISM which has a number density $N_\mathrm{H}=1$~cm$^{-3}$
typical for the galactic disk, the primary GCR reacceleration is not
important.  Only in the case of a rarefied ISM, like a hot ISM with a
number density $N_\mathrm{H}=0.003$~cm$^{-3}$, one can expect an
important contribution of GCR reacceleration to their overall spectrum
$n_\mathrm{GCR}(\epsilon_\mathrm{k})$. The same conclusion has been
reached for CR acceleration in the ISM of elliptical galaxies that has
similar parameters (Dorfi \& V\"olk \cite{dv96}).

Due to the additional factor $(p/p_\mathrm{GCR})^{\mu}$ the
contribution of reacceleration becomes dominant in the secondary GCR
spectra at high energies. The value of the critical energy above which
the secondary spectra are expected to be as hard as the primary ones
depends upon the mean gas density $N_\mathrm{g}^\mathrm{SCR}$ inside
the SNRs: according to the above formulae, it is expected to be
100~GeV/n for $N_\mathrm{g}^\mathrm{SCR}\sim 10^{-2}$~cm$^{-3}$, and an
order of magnitude larger if $N_\mathrm{g}^\mathrm{SCR}\sim
1$~cm$^{-3}$.

There is an additional mechanism of secondary GCR production inside SNRs:  
primary nuclei like GCRs in the Galactic disk produce light secondary as a
result of nuclear spallation due to their interaction with the background
gas. One can easily estimate the relative significance of this mechanism
compared with the previous one. The total production rate of secondaries
in SNRs due to the spallation of primaries is
\begin{equation}
Q_\mathrm{s}^\mathrm{SCR}\propto
N_\mathrm{g}^\mathrm{SCR}N_\mathrm{SN}N_\mathrm{p},
\label{eq19}
\end{equation}
where $N_\mathrm{SN}$ is the number of existing SNRs in the Galaxy and 
$N_\mathrm{g}^\mathrm{SCR}$ is
mean number density of the thermal gas in SNRs. GCRs generate the
same kind of secondaries with the rate
\begin{equation}
Q_\mathrm{s}^\mathrm{GCR}\propto
N_\mathrm{g}^\mathrm{GCR}V_\mathrm{g} n_\mathrm{p},
\label{eq20}
\end{equation}
where $N_\mathrm{g}^\mathrm{GCR}\approx 1$~cm$^{-3}$ is mean gas number 
density in the
Galactic disk. Taking into account that the SNR number
$N_\mathrm{SN}=T_\mathrm{SN}\nu_\mathrm{SN}$ is determined by the CR 
confinement time $T_\mathrm{SN}$
inside a SNR, and that the GCR number density $n_\mathrm{p}$ is related to the
overall CR spectrum produced in SNR $N_\mathrm{p}$ by the relation
$n_\mathrm{p}=N_\mathrm{p} \nu_\mathrm{SN} \tau_\mathrm{g}/V_\mathrm{g}$, we have
\begin{equation}
\frac{Q_\mathrm{s}^\mathrm{SCR}} {Q_\mathrm{s}^\mathrm{GCR}}=
\frac{N_\mathrm{g}^\mathrm{SCR}}{N_\mathrm{g}^\mathrm{GCR}}
\frac {T_\mathrm{SN}}{\tau_\mathrm{g}}.
\label{eq21}
\end{equation}
Like reacceleration, this mechanism produces a substantially harder spectrum
of secondaries compared with the spectrum created by the GCRs in the
Galactic disk. It has the opposite dependence on the ISM density: it is
more efficient in a denser ISM.

Since $T_\mathrm{SN}\sim 10^5$~yr and since $\tau_\mathrm{g}\approx 5\times 10^6$~yr at
GeV energies (e.g. Berezinskii et al. \cite{brz90}), we have
$Q_\mathrm{s}^\mathrm{SCR}/Q_\mathrm{s}^\mathrm{GCR}=2\times 10^{-2}$ 
if on average SNe explode into the
ISM with a gas number density $N_\mathrm{g}^\mathrm{SCR}=
N_\mathrm{g}^\mathrm{GCR}$. This value is
somewhat lower than the corresponding contribution of reacceleration to
the production of secondary nuclei. At the same time, the spectrum of
secondaries spectrum $N_\mathrm{s}(p)$ has in this case exactly the same form as
that of the primaries, $N_\mathrm{p}(p)$. As demonstrated below, in the case of
reacceleration the ratio $N_\mathrm{s}(p)/N_\mathrm{p}(p)$ 
decreases slightly with momentum
$p$ due to the fact that secondaries are more effectively produced at the
late SNR evolutionary phase when the SN shock becomes weaker, whereas high
energy primaries are mainly accelerated at the beginning of the Sedov
phase, when the SN shock is extremely strong. Therefore one can expect
that the secondary nuclei production due to the hadronic SCR collisions
with the gas nuclei makes a considerable contribution at high energies
$\epsilon_\mathrm{k}\gsim 100$~GeV/n.

The production of secondaries due to spallation of primary SCRs in SNRs is
described in our model by the source term
\begin{equation}
q_\mathrm{s}(r,p,t)=\sigma_\mathrm{ps}vN_\mathrm{g}f_\mathrm{p}(r,p,t)
\label{eq22}
\end{equation}
in the transport equation for the distribution function
$f_\mathrm{s}(r,p,t)$ of secondary nuclei. Here $f_\mathrm{p}(r,p,t)$ is
the distribution function of primary nuclei and $N_\mathrm{g}$ is the
local gas number density, both of which are also functions of SNR
evolutionary time and position. The distribution function $f(r,p,t)$,
which determines all relevant CR characteristics, is calculated in our
model selfconsistently for all CR elements considered (see Berezhko et al.
\cite{byk96} for details).

The actual situation with the CR spectra in SNRs is rather more
complicated than suggested by the above estimates. Due to nonlinear
acceleration effects, CR spectra produced in SNRs are not pure power laws.
Besides that particles with different energies are effectively produced
during different phases of SNR evolution, whereas the above simple
estimates suggest in fact that the entire overall CR spectrum $N(p)$ is
produced at an epoch characterized by the SNR volume $V_\mathrm{SN}$ and age
$T_\mathrm{SN}$. To reach more definitive conclusions about the role of CR
reacceleration in SNRs, we need to perform a selfconsistent consideration.
This can only be done numerically.

Note that the source CR spectra $N(\epsilon_\mathrm{k})$ which are the time
integral spectra of the CRs accelerated during the whole active period of
SNR evolution, depend on the GCR spectra $n(\epsilon_\mathrm{k})$ since not only
the injection of suprathermal particles leads to the formation of
$N(\epsilon_\mathrm{k})$ but the injection and the reacceleration of existing
background GCRs contribute as well. This means that the spectra
$N(\epsilon_\mathrm{k})$ should be determined as the result of a selfconsistent
solution of the equations which describe the GCR propagation and nuclear
transformation in the Galactic confinement volume, together with the
equations, which govern SNR evolution and CR acceleration. Fortunately
this rather complicated problem can be considerably simplified without
loss of accuracy. Indeed, as discussed before, 
due to the steep GCR spectra at relativistic
energies $\epsilon_\mathrm{k}\gg 1$~GeV/n, only the injection of 
GCRs with energies
$\epsilon_\mathrm{k} \lsim 1$~GeV/n plays a role for the formation of 
the resultant
spectra $N(\epsilon_\mathrm{k})$. Since the GCR spectra 
$n(\epsilon_\mathrm{k})$ at these
subrelativistic energies are known from experiment, we use them as
physical input like other ISM parameters in our selfconsistent nonlinear
kinetic model for CR acceleration in SNR, so that the output of this
model, which is the overall CR spectrum $N(\epsilon_\mathrm{k})$, can be
considered as consistently determined together with the GCR spectra
$n(\epsilon_\mathrm{k})$.

\begin{figure}
\centering
\includegraphics[width=7.5cm]{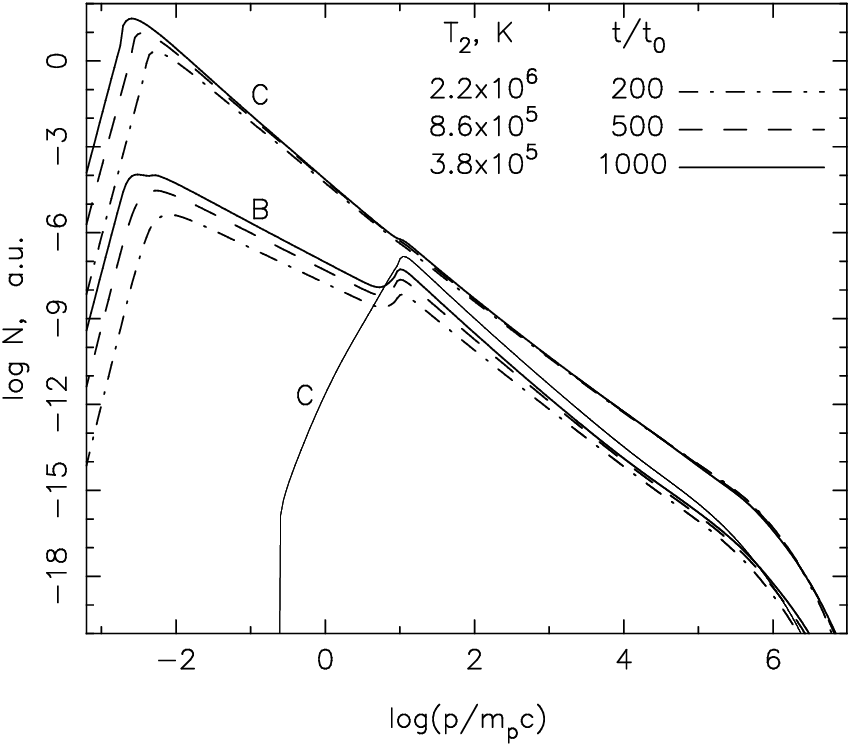}
\caption{The overall momentum spectra of B and C nuclei produced during
the SNR evolution in the warm ISM ($N_\mathrm{H}=0.3$~cm$^{-3}$, $T=10^4$~K,
$t_0=367$~yr)  up to three different moments of time.  The carbon spectra,
starting from $p/m_\mathrm{p}c\ll 1 $, are due to the acceleration of injected
suprathermal particles and GCR reacceleration, whereas the boron spectra
are the result of reacceleration of GCRs and the spallation of carbon and
oxygen nuclei inside SNR.  For clarity, the spectrum of purely
reaccelerated GCR carbon nuclei, corresponding to $t/t_0=1000$, is shown
by the thin solid line.}
\label{fig1}
\end{figure}

\section{Results and discussion}

In order to study the effect of GCR reacceleration in SNRs on the s/p
ratio we consider the elements boron (B) and carbon (C) because carbon is
the most abundant element in the GCRs and plays a major role in the
production of boron which is usually considered as a pure secondary. We
have performed selfconsistent calculations of CR acceleration in SNRs,
based on the kinetic nonlinear model, for the simple case of a uniform ISM
with different densities.

We use the values $E_\mathrm{SN}=10^{51}$~erg for the explosion energy and
$M_\mathrm{ej}=1.4 M_{\odot}$ for the ejecta mass which are typical for SNe~Ia in
a uniform ISM. Note that the main fraction of the core collapse SNe has
relatively small initial progenitor star masses between 8 and 15
$M_{\odot}$ which therefore do not significantly modify the surrounding
ISM through the main sequence wind of the progenitor star (e.g. Abbott
\cite{abbott}). SNR evolution in this case is very similar to that of SNe~Ia.

Only in the case of type Ib and type II SNe with massive progenitor stars
$M_\mathrm{i}>15M_{\odot}$, SNRs expand into a nonuniform circumstellar medium
strongly modified by the intense progenitor star wind. The SNR evolution
looks differently compared to the case of a uniform ISM. At the same time,
the main amount of CRs is expected to be produced when the SN shock
propagates through the hot rarefied bubble, where the evolution and CR
acceleration are roughly similar to that in the case of a uniform ISM
which has the same parameters as the bubble (Berezhko and V\"olk
\cite{bv00}).
Therefore the contribution of SNRs expanding in the strongly modified
circumstellar medium effectively increase the filling factor of the hot
phase of ISM.

We adopt an injection rate of suprathermal particles,
characterized by the injection parameters $\eta=10^{-4}$ and $\lambda=4$,
which are expected for a typical SNR (V\"olk et al. \cite{vbk03}).
GCR nuclei B and C with number densities $N_\mathrm{B}=7.9\times
10^{-14}$~cm$^{-3}$ and $N_\mathrm{C}= 2.6\times 10^{-13}$~cm$^{-3}$, respectively,
are injected at a kinetic energy $\epsilon_\mathrm{inj}=0.6$~GeV/n which
corresponds to the mean GCR energy for these elements. We consider three
essentially different phases of the ISM: a diluted, hot ISM with hydrogen
number density $N_\mathrm{H}=0.003$~cm$^{-3}$ and temperature $T_0=10^6$~K, a warm
ISM with $N_\mathrm{H}=0.3$~cm$^{-3}$ and $T_0=10^4$~K, and an "average" ISM with
$N_\mathrm{H}=1$~cm$^{-3}$ and $T_0=10^4$~K. The ISM magnetic field values
$B_0=3$~$\mu$G, $5$~$\mu$G and 10~$\mu$G where taken for these three
cases, respectively.

\begin{figure}
\centering
\includegraphics[width=7.5cm]{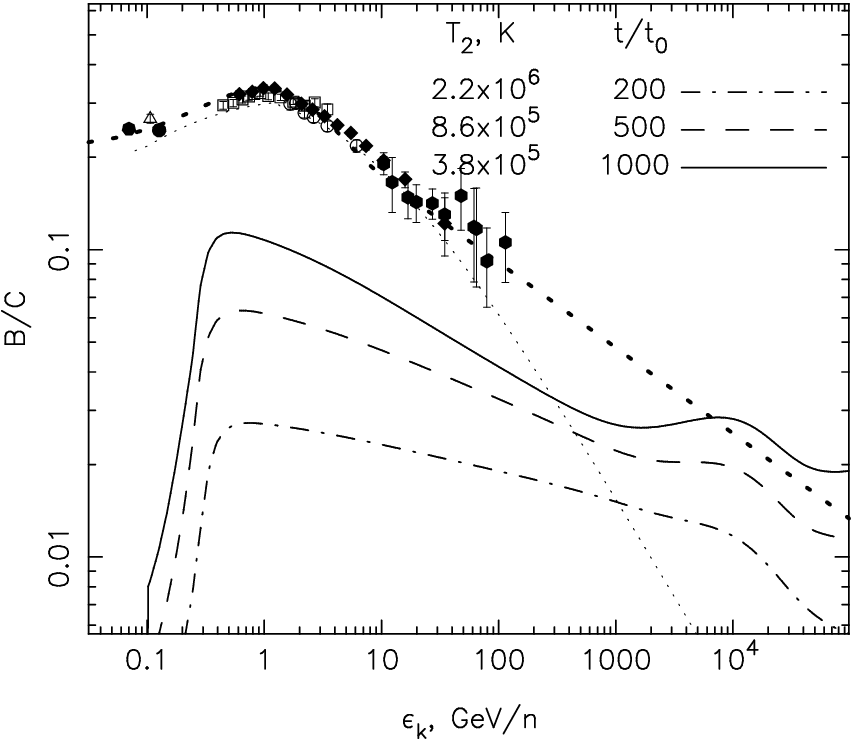}
\caption{The B/C ratio $r_\mathrm{g}=n_\mathrm{B}/n_\mathrm{C}$ as a 
function of kinetic energy per
nucleus. {\it Solid, dashed and dash-dotted lines} represent the ratio
$r_\mathrm{s}=N_\mathrm{B}/N_\mathrm{C}$ for the case presented in 
Fig.~\ref{fig1}. {\it Dotted lines}
represent the fit of the experimental data within the leaky box model
({\it thin dots}) on the one hand, and within the diffusive model with
distributed reacceleration in the ISM ({\it heavy dots}) on the other.
Experimental data (Engelmann et al. \cite{eng90}) are also shown.}
\label{fig2}
\end{figure}

Numerical results, which correspond to the so-called warm ISM with
$N_\mathrm{H}=0.3$~cm$^{-3}$, are presented in Figs.~\ref{fig1} and \ref{fig2}. The 
overall momentum spectra of boron and carbon
\begin{equation}
N(p,t)=16\pi^2 p^2\int_0^{\infty}dr r^2 f(r,p,t),
\label{eq23}
\end{equation}
which include all particles accelerated during the SNR evolution up to
time $t$, are shown in Fig.~\ref{fig1} for three successive times $t$. One can see
that reacceleration of GCRs makes quite a small contribution to the
overall spectrum of carbon, in agreement with the earlier estimate.
Note also that the late evolutionary phases are not very important for the
highest energy part of the spectrum of primary nuclei: the C-spectrum at
$p>10^3m_\mathrm{p}c$ is formed mainly at the beginning of the Sedov phase ($t\sim
t_0$), and it is almost unchanged at late phases $t>100t_0$. There are two
physical factors which determine the above situation. The first is a
geometrical factor which results in the most efficient acceleration of the
highest energy CRs at the beginning of the Sedov phase
(Berezhko \cite{ber96}). Due to the
decrease of the product $R_\mathrm{s}V_\mathrm{s}$, the maximum energy of freshly
injected/accelerated CRs decreases with time in the Sedov phase. Therefore
the late Sedov phases do not contribute significantly to the highest
energy part of the resulting CR spectrum.

The second factor is the decrease of the injected suprathermal particle
momentum $p_\mathrm{inj}\propto V_\mathrm{s}$ due to the shock deceleration. 
It diminishes
the contribution of the late phases to CR production. As a result, the
overall spectrum of CR primary nuclei is mainly formed during the early
Sedov phase when the shock is extremely strong, and therefore it is quite
hard.

In the case of secondaries the situation is very different. In the
B-spectra two different components are clearly visible. The first one,
peaking at $p=p_\mathrm{GCR}\approx 10 m_\mathrm{p}c$, 
is due to the injection of GCRs with momentum
$p_\mathrm{GCR}$.  Since $p_\mathrm{GCR}$ does not depend on time, 
the efficiency of
their acceleration (reacceleration) progressively increases with time,
roughly proportional to the total number of particles participating in the
acceleration. This number is proportional to the shock volume $V_\mathrm{SN}$.
Since in this case the late SN evolutionary phases, when the SN shock is
no more so strong as during the early Sedov phase, are important for
secondary CR production, their overall spectrum is much steeper than the
spectrum of primary nuclei (see Fig.~\ref{fig1}).

The second component in the B-spectra is due to the spallation of carbon
and oxygen nuclei. Its spectrum is substantially harder than that of the
first component, because it is directly related to the spectra of
primaries. Therefore it contributes strongly at very high momenta
$p\gsim10^4m_\mathrm{p}c$ and also dominates at very low momenta 
$p<p_\mathrm{GCR}$, where
B production due to reacceleration is negligible.

On account of the above factors the ratio $r_\mathrm{s}$, which is exclusively due
to the production of secondaries in SNRs, becomes steeper and increases
with time, as one can see from Fig.~\ref{fig2}. Therefore the effect of GCR
reacceleration on the s/p ratio depends on the confinement time $T_\mathrm{SN}$
which corresponds to the age of the SNR when the shock becomes an
inefficient accelerator and release of previously accelerated CRs into the
Galactic volume proceeds.

There are at least two conditions which determine the confinement time
$T_\mathrm{SN}$.

The first one is the decrease of the shock Mach number during the SNR
evolution. When it becomes so low, at some stage $t=t_4$, such that the
shock compression ratio $\sigma$ drops below 4, the acceleration of
freshly injected particles becomes inefficient. This leads to the decrease
of the turbulence level near the shock front and to subsequent escape of
previously accelerated CRs due to increase of their diffusive mobility.
This factor plays the major role in the case of a hot ISM.

In fact, the so-called effective compression ratio 
\begin{equation}
\sigma_\mathrm{ef}=\sigma(1-1/M_\mathrm{a}),
\label{eq24}
\end{equation}
rather than $\sigma$, plays the main role in CR acceleration and their
final spectrum. Due to the fact, that Alfv\'en waves, excited by the
accelerated CRs upstream of the shock, propagate in upstream direction
with Alfv\'en speed $c_\mathrm{a}$, the effective compression ratio of scattering
centers seen by CRs $\sigma_\mathrm{ef}$ is always smaller than $\sigma$. This
effect is described in the above expression by the Alfv\'en Mach number
$M_\mathrm{a}=c_\mathrm{a} /V_\mathrm{s}$. Therefore $t_4$ is the time 
when $\sigma_\mathrm{ef}$ drops below 4.

The acceleration process may also stop at some stage $t=t_6$ when the
postshock temperature drops below $10^6$~K, because radiative SNR cooling
sets in strongly. Since the SNR looses a large amount of its energy,
efficient CR acceleration terminates presumably at this stage.
 
We adopt here a confinement time, which is the minimum of these two:
\begin{equation}
T_\mathrm{SN}=min\{t_4, t_6\}.
\label{eq25}
\end{equation}
In the case of a warm ISM, we have a postshock gas temperature
$T_2=1.3\times 10^6$~K at $t=300t_0$, and therefore the confinement time
is $T_\mathrm{SN}=t_6\approx 10^5$~yr ($t_0=367$~yr in this case). The relevant 
ratio $r_\mathrm{s}$ in Fig.~\ref{fig2} is therefore given by the dashed line.


The significance of GCR reacceleration can be understood if one compares
the expected source s/p ratio $r_\mathrm{s}$ with the ratio $r_\mathrm{g}$, 
extracted from
the measured B/C ratio in the standard model.
 
Since the boron nuclei are produced in the ISM not only by spallation of
carbon but also by spallation of oxygen, the expression (\ref{eq4}) has to be
rewritten in the form:
\begin{equation}
\frac{n_\mathrm{B}}{n_\mathrm{C}}=
\frac{x}{m_\mathrm{p}+\sigma_\mathrm{B} x}
\left(\sigma_\mathrm{CB}
+\sigma_\mathrm{OB}\frac{n_\mathrm{O}}{n_\mathrm{C}}\right)+
\frac{N_\mathrm{B}}{N_\mathrm{C}}
\frac{(m_\mathrm{p}+\sigma_\mathrm{C} x)}{(m_\mathrm{p}+\sigma_\mathrm{B} x)}.
\label{eq26}
\end{equation}

This expression is used in order to determine the expected B/C ratio
with accounts for boron production inside the GCR sources (in SNRs).
In the case when the secondaries are not produced in CR sources we have
from this expression
\begin{equation}
r_\mathrm{g}=\frac{(\sigma_\mathrm{CB}+\sigma_\mathrm{OB})x_\mathrm{g}}
{m_\mathrm{p}+\sigma_\mathrm{B} x_\mathrm{g}}.
\label{eq27}
\end{equation}

According to the standard leaky box model, the value of the escape length
$x_\mathrm{g}\approx 14$~g/cm$^2$ at GeV energies with energy dependence
$x_\mathrm{g}\propto \epsilon_\mathrm{k}^{-0.6}$ at $\epsilon_\mathrm{k}>5$~GeV/n fits the
existing B/C data (Engelmann et al. \cite{eng90}).

Note however that the value of the escape length derived from the same
experimental data depends also on the GCR propagation model, especially at
high energies where the existing experimental data do not constrain it.

The galactic model with distributed stochastic reacceleration predicts a
different (weaker) decrease of the B/C ratio at high energies $\epsilon_\mathrm{k}
> 30$~ GeV/n than does the standard leaky box model. According to Jones et
al. (\cite{jlpw01}), $x_\mathrm{g}\propto\epsilon_\mathrm{k}^{-0.54}$ in the model without
reacceleration, and $x_\mathrm{g}\propto \epsilon_\mathrm{k}^{-0.3}$ in the model with
stochastic reacceleration. Both models give the same s/p ratio at
$\epsilon_\mathrm{k} < 30$~GeV/n.
\begin{figure}
\centering
\includegraphics[width=7.5cm]{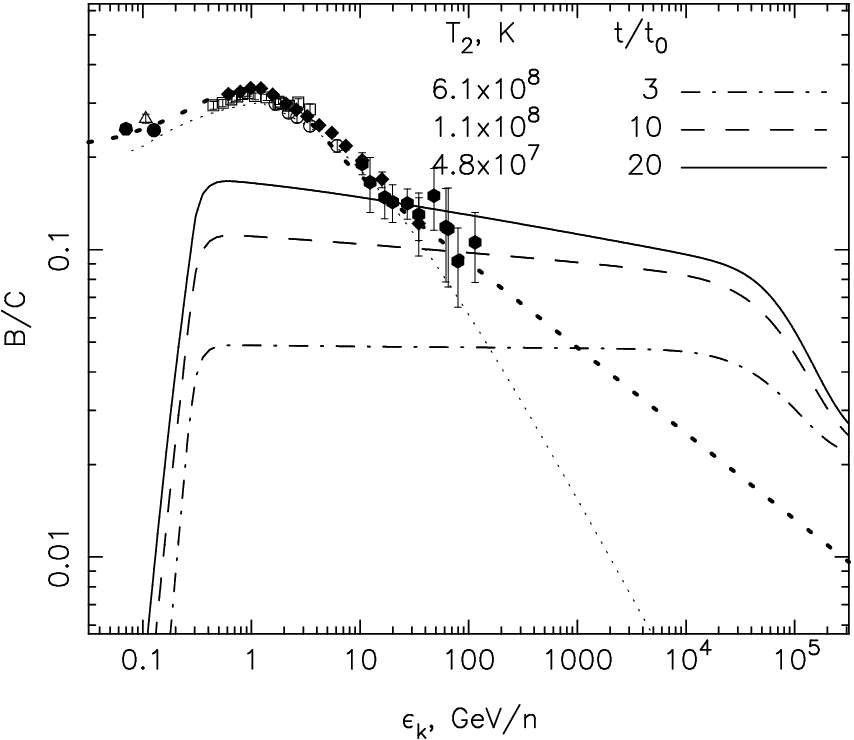}
\caption{The same as in Fig.~\ref{fig2} but for the case of a hot ISM 
($N_\mathrm{H}=0.003$~cm$^{-3}$, $t_0=1674$~yr).}
\label{fig3}
\end{figure}

\begin{figure}
\centering
\includegraphics[width=7.5cm]{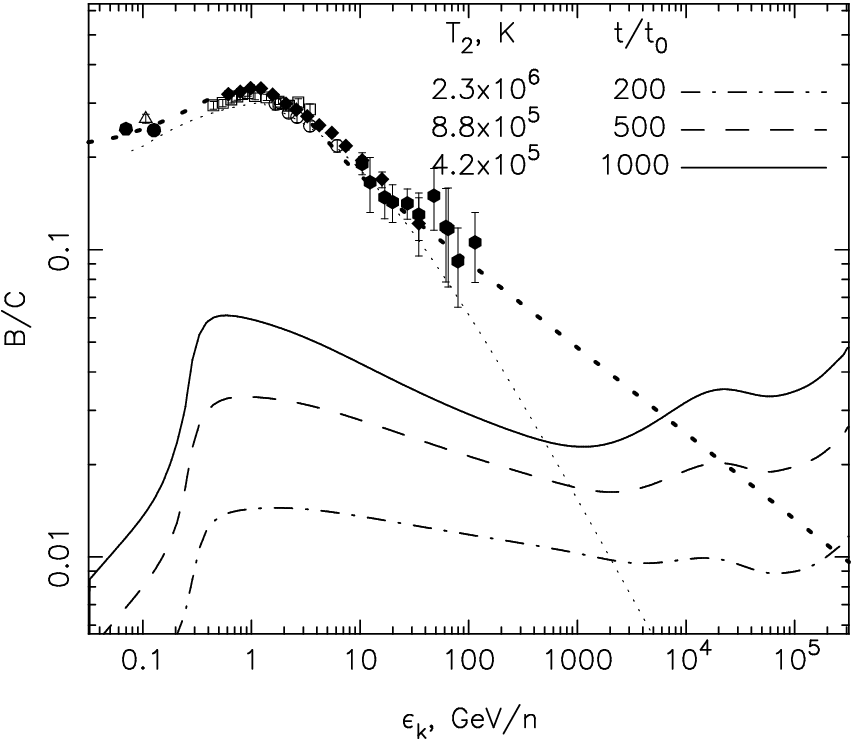}
\caption{The same as in Fig.~\ref{fig2} but for the case of an ISM hydrogen number 
density
$N_\mathrm{H}=1$~cm$^{-3}$ ($t_0=246$~yr).}
\label{fig4}
\end{figure}

\begin{figure}
\centering
\includegraphics[width=7.5cm]{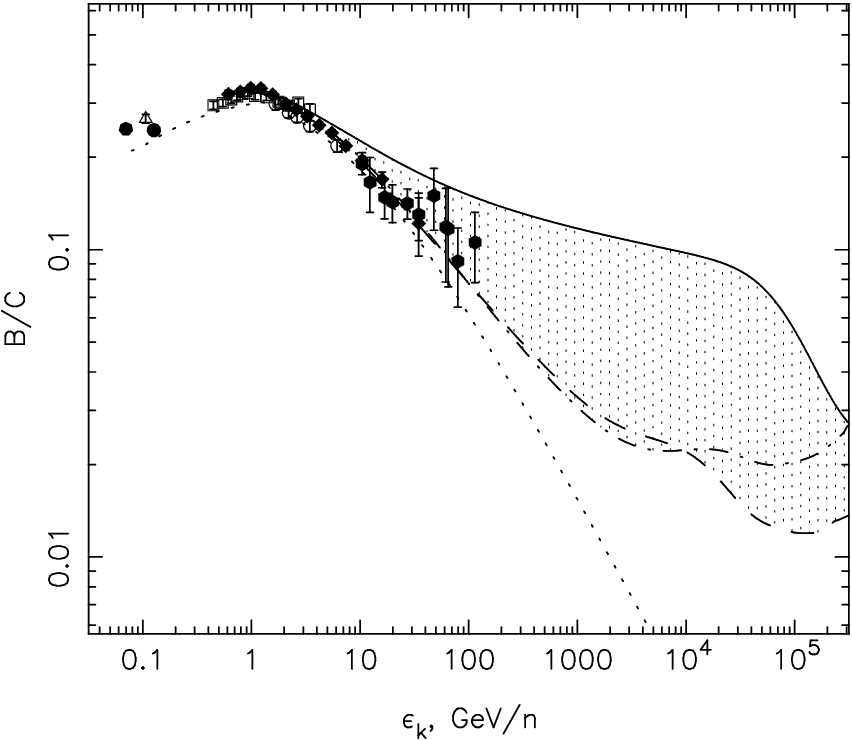}
\caption{The expected overall B/C ratio as a function of kinetic energy
per nucleus. The three curves, corresponding to the three circumstellar
hydrogen number densities $N_\mathrm{H}=0.003$~cm$^{-3}$ ({\it solid line}),
$N_\mathrm{H}=0.3$~cm$^{-3}$ ({\it dashed line)}, and $N_\mathrm{H}=1$~cm$^{-3}$ ({\it
dash-dotted line}) encompass a range given by the {\it dotted area}.}
\label{fig5}
\end{figure}

\begin{figure}
\centering
\includegraphics[width=7.5cm]{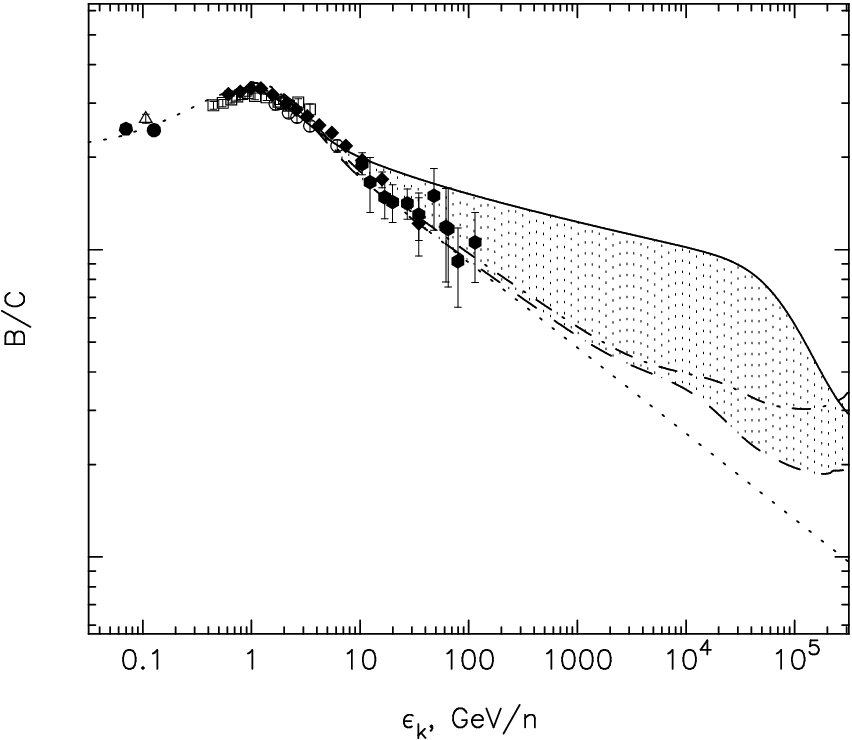}
\caption{The same as in Fig.~\ref{fig5} but comparing with the mean CR escape length
  $x_\mathrm{g}(\epsilon_\mathrm{k})$ predicted by the diffusion model with 
  distributed CR reacceleration.}
\label{fig6}
\end{figure}

In order to demonstrate how significantly secondary CR production in SNRs
changes the expected s/p ratio, we consider here two extreme predictions
for $x_\mathrm{g}\propto \epsilon_\mathrm{k}^{-\mu}$. The first one, with $\mu=0.6$,
corresponds to the standard leaky box model (Engelmann et al. \cite{eng90}). The
second, with $\mu=0.3$, corresponds to the diffusion model with
distributed stochastic reacceleration (Jones et al. \cite{jlpw01}). The B/C ratio
calculated with these two functions $x_\mathrm{g}(\epsilon_\mathrm{k})$ according to the
expression (\ref{eq26}) with $N_\mathrm{B}=0$, are shown in
Figs.~\ref{fig2}--\ref{fig6} by the dotted lines.
The cross section values and the ratio $n_\mathrm{O}/n_\mathrm{C}$, required to calculate
the B/C ratio, are taken from the paper by Engelmann et al. (\cite{eng90}).
 
In Figs.~\ref{fig2}--\ref{fig4} we also present the ratio
$r_\mathrm{s}=N_\mathrm{B}/N_\mathrm{C}$ of the boron and carbon
spectra, produced in SNRs by the combination of all processes considered 
up to different times $t/t_0$ and
calculated for three different ISM densities. One can see from
Fig.~\ref{fig2},
which corresponds to $N_\mathrm{H}=0.3$~cm$^{-3}$, that the source ratio $r_\mathrm{s}$
exceeds the B/C ratio $r_\mathrm{g}$ predicted by the leaky box model at an energy
$\epsilon_\mathrm{k}>300$~GeV/n and only at $\epsilon_\mathrm{k}>10$~TeV/n it 
 becomes larger than $r_\mathrm{g}$ which corresponds to the diffusive model with
distributed reacceleration if the CR confinement time is as large as
$T_\mathrm{SN}=4\times 10^5$~yr. Note however, that at such a late time
$t=10^3t_0=4\times 10^5$~yr, the postshock temperature is already
$T_2\approx 4\times 10^5$~K, and in addition the effective compression
ratio drops up to $\sigma_\mathrm{ef}=3.25$. Therefore the CR confinement time is
taken to be equal $T_\mathrm{SN}=500t_0=2\times 10^5$~yr, because the postshock
temperature at this stage is $T_2= 9\times 10^5$~K and the effective
compression ratio has a value $\sigma_\mathrm{ef}=3.8$ which is only slightly
below 4.

In Fig.~\ref{fig3} we present the same as in Fig.~\ref{fig2} but for the rarefied, so-called
hot ISM which is characterized by a temperature $T_0=10^6$~K and a hydrogen
number density $N_\mathrm{H}=0.003$~cm$^{-3}$. One can see that the ratio
$r_\mathrm{s}$ exceeds at energies $\epsilon_\mathrm{k} \gsim 100$~GeV/n the
measured value of B/C ratio already at $t=3\times 10^4$~yr.

We note that already at this stage the effective compression ratio drops
below 4 and the acceleration of freshly injected particles becomes
progressively less efficient. Therefore one can assume that in this case
$T_\mathrm{SN}=t_4\approx 3\times 10^4$~yr. Even for this relatively short
confinement time the contribution of the reacceleration process in the B/C
ratio is strong at energies $\epsilon_\mathrm{k}$ greater than 10~GeV/n (solid
curve in Fig.~\ref{fig3}). Note that at such a low ISM density the production of
secondary nuclei in SNRs is entirely due to their reacceleration.

At larger ISM densities secondary nuclei production due to the primary CR
spallation becomes more important. This is illustrated in Fig.~\ref{fig4}, where the
expected B/C ratio $r_\mathrm{s}$ is represented for $N_\mathrm{H}=1$~cm$^{-3}$. It is seen
that at high energies $\epsilon_\mathrm{k}\gsim 10^4$~GeV/n the expected ratio
$r_\mathrm{s}$ is higher than in the case $N_\mathrm{H}=0.3$~cm$^{-3}$. Therefore, we
conclude that the lowest efficiency of secondary nuclei production in SNRs
occurs in the case of ISM number density $N_\mathrm{H}=0.3$~cm$^{-3}$. It becomes
larger both in a denser and a more diluted ISM because of efficient
primary spallation and GCR reacceleration, respectively. Since at
$t=500t_0$ the postshock temperature $T_2=9\times 10^5$~K is already lower
than $10^6$~K, the CR confinement time is $T_\mathrm{SN}=t_6 \approx 10^5$~yr in
this case even though the effective shock compression ratio
$\sigma_\mathrm{ef}=4.2$ is still slightly higher than 4.

Since the typical ISM density in which the majority of SN
explosions occurs is not well known, we present in Figs.~\ref{fig5} and
\ref{fig6} the
possible range of the expected overall B/C ratios, calculated according to 
expression (\ref{eq26}).
The source ratios $r_\mathrm{s}=N_\mathrm{B}/N_\mathrm{C}$ are taken from 
the above calculations and
the CR escape length
\begin{equation}
x=\delta x_\mathrm{g}
\label{eq28}
\end{equation}
was assumed to be some fraction $\delta<1$ of the escape length
\begin{equation}
x_\mathrm{g}=\frac{m_\mathrm{p}r_\mathrm{g}}
{r_\mathrm{g}\sigma_\mathrm{B}-\sigma_\mathrm{CB}-\sigma_\mathrm{OB}},
\label{eq29}
\end{equation}
determined from the relation (\ref{eq27}). The cross section values and the ratio
$n_\mathrm{O}/n_\mathrm{C}$, required to calculate the B/C ratio, are taken from the paper 
by
Engelmann et al. (\cite{eng90}). To fit the existing data one needs $\delta=0.3$,
0.7 and 0.9 for the ISM number densities $N_\mathrm{H}=0.003$~cm$^{-3}$,
0.3~cm$^{-3}$ and 1~cm$^{-3}$, respectively. The B/C ratio is shown 
for these three densities. The production of
secondary CRs in SNRs leads to a substantial decrease of the escape length
$x$ only if all SNRs explode into a diluted ISM with hydrogen number
density $N_\mathrm{H}=0.003$~cm$^{-3}$. In the two other cases, which are expected
to be more probable, the relative decrease of $x$ is not so important.
It is significantly lower compared with the estimates of Wandel
(\cite{wan88}),
because we believe that the active period of SNR evolution stops much
earlier than he assumed.
%

It is seen that for $N_\mathrm{H}=0.003$~cm$^{-3}$ the boron production in SNRs
leads to considerably higher values of the B/C ratio at energies
$\epsilon_\mathrm{k}>100$~GeV/n than the observations show. There are however some
experimental indications that the measured B/C ratio becomes indeed
flatter at energies near 100~GeV/n. It is interesting to see that the
theoretical curves do not change monotonically with gas density. In fact a
weighted mixture of SN sites with very low densities,
$N_\mathrm{H}=0.003$~cm$^{-3}$ (hot ISM), and moderate densities 
$N_\mathrm{H}=0.3$~cm$^{-3}$
(warm ISM) is able to approximate the observations.
More precise measurements at these energies are needed, in order to draw
definite conclusions about the actual role of GCR reacceleration.

We note that the wide spread of the B/C ratio at high energies is due to
the unknown correlations of SNR sites with gas density. Nevertheless it is
clear from Figs.~\ref{fig2}--\ref{fig4}, that at any given density the 
energy dependence of
the B/C ratio is expected to be much flatter than predicted by the leaky
box model.

We have a similar situation in the case when the GCR escape length
$x_\mathrm{g}(\epsilon_\mathrm{k})$, predicted by the diffusive model with
distributed reacceleration, is used for comparison with the theory, as one
can see from Fig.~\ref{fig6} for the same three densities. Since the
energy dependence of the ratio $r_\mathrm{g}(\epsilon_\mathrm{k})$ is much
flatter in this case, the relative increase of the B/C ratio due secondary
CR production in SNRs is much smaller than in the previous case.
Nevertheless, the expected value of the B/C ratio for
$\epsilon_\mathrm{k}\gsim 1$~TeV/n is still noticeably higher, by a factor
of two on average, than predicted by the model which neglects secondary CR
production in SNRs. The CR escape length values are determined by the
factor $\delta=0.2$, 0.6 and 0.8 for $N_\mathrm{H}=0.003$~cm$^{-3}$,
0.3~cm$^{-3}$ and 1~cm$^{-3}$, respectively.

One could expect that we deal with a more realistic situation when a 
variety of existing SNRs in the Galaxy are expanding into ISM regions
with different  densities. In order to make a definite prediction about
the expected B/C ratio one  should determine properly weighted boron
and carbon spectra:
\begin{equation}
N_{B,C}(\epsilon_k)=\Sigma \Delta \nu_{SN}^i 
N_{B,C}(\epsilon_k, N_g^i)/\nu_{SN}
\end{equation}
for a given SN explosion rate $\Delta \nu_{SN}^i$ in an ISM with number
density $N_g^i$. Unfortunately the explosion rates $\Delta \nu_{SN}^i$ are
not known. At the same time, as it is clear from Figs. 5 and 6, the
uncertainty in the B/C ratio is determined mainly by the value $\Delta
\nu_{SN}^h$, which corresponds to the hot ISM. Since the boron spectrum is
much more sensitive to the ISM density than the carbon spectrum, we can
approximately write
\begin{equation}
B/C=(B/C)_h\Delta \nu_{SN}^h/\nu_{SN} +(B/C)_w(1-\Delta \nu_{SN}^h/\nu_{SN}),
\end{equation}
taking into account that within the ISM density range from 0.3 to
1~cm$^{-3}$ the ratio B/C changes not so significantly.  Here $(B/C)_h$
and $(B/C)_w$ are the B/C ratios produced in hot and warm ISM,
respectively. This formula makes it possible to estimate the expected B/C
ratio for a given value $\Delta \nu_{SN}^h/\nu_{SN}$ without making
additional calculations.

\section{Summary}

Our considerations based on the selfconsistent kinetic nonlinear model for
CR acceleration in SNRs show that the reacceleration of the existing GCRs
and the spallation of primary CRs due to their collisions with the gas
nuclei in SNRs strongly influence the energy spectra of secondary
elements, like Li, Be, B. Due to this additional mechanism the spectra of
secondaries become significantly flatter at high energies $\epsilon_\mathrm{k} >
100$~GeV/n.  This effect can be directly studied from the s/p ratio: at
high energies where reacceleration is important, the s/p ratio should be
flatter than at lower energies $1\lsim \epsilon_\mathrm{k} \lsim 100$~GeV/n.

Compared with previous considerations based on the plane-wave approach
(Wandel \cite{wan88}), our model selfconsistently includes a number of
physical factors: i) nonlinear shock modification due to the CR
backreaction; ii) CR adiabatic cooling inside the expanding SNR
interior;   iii) the so-called CR escape effect. In addition, all
stages of SNR evolution contribute to the resulting overall CR spectra,
in contrast to the assumption by Wandel that only very late SNR phases
play a role.

Qualitatively, the efficiency of secondary CR nuclei production in SNRs
depends in an important way on the density of the ISM in which SNRs are
exploding. The lower the ISM density is, the larger is the volume of the
SNR which is reached during the active period of its evolution.
Correspondingly larger numbers of background GCRs are swept-up and
reaccelerated compared with the number of particles injected from the
thermal postshock distribution. If SNRs are on average exploding into a
diluted ISM, then the influence of reacceleration becomes significant
already at $\epsilon_\mathrm{k} \sim 10$~GeV/n. This is in contradiction to the
existing data for the B/C ratio and means that the typical ISM density
"seen" by SNe is greater than 0.003~cm$^{-3}$.

An increase of the ISM density leads to a decrease of the reacceleration
effect. At the same time the production of secondaries in SNRs due to the
spallation of primaries becomes more important. If the typical density is
between about $N_\mathrm{H}=0.3$ and 1~cm$^{-3}$, reacceleration and spallation
produce roughly equal contributions to the B/C ratio at $\epsilon_\mathrm{k}\approx
1$~TeV/n, whereas at higher energies spallation becomes dominant.
Quantitatively the effect depends upon the mean GCR escape length and its
energy dependence. In the case $N_\mathrm{H}=0.3$~cm$^{-3}$ boron production in
SNRs becomes dominant for $\epsilon_\mathrm{k}>0.5$~TeV/n within the leaky box
model, and comparable within the diffusion model with
distributed reacceleration for $\epsilon_\mathrm{k}>3$~TeV/n. 

The value of the CR confinement time $T_{SN}$ inside SNRs plays the
crucial role for the secondary CR production inside SNRs: the larger
$T_{SN}$ is, the greater is the number of secondaries that are produced.  
In the case of a diluted ISM with number density $N_H=0.003$~cm$^{-3}$, CR
confinement is presumably restricted to times smaller than $t_4$ when the
shock becomes too weak to support a high level of MHD turbulence near the
shock front due to the CR streaming instability. In the case of a much
denser ISM with $N_H\sim 1$~cm$^{-3}$ the active period of CR production
in SNRs is assumed to terminate when the postshock temperature drops below
$10^6$~K. For intermediate ISM densities $N_H\sim 0.3$~cm$^{-3}$ both
effects are equally important. Due to these two physical factors the CR
confinement time $T_{SN}$ is significantly lower in our model compared
with previous considerations (Wandel \cite{wan88}), where these factors
were ignored. This is one of the reasons why the effect of secondary CR
production in SNRs is significantly lower in our model compared with the
prediction of Wandel \cite{wan88}.

If on average SNe explode in the ISM with $N_\mathrm{H}=1$~cm$^{-3}$ boron
production in SNRs leads to an extremely flat, almost energy-independent
B/C ratio at
$\epsilon_\mathrm{k}>1$~TeV/n and at $\epsilon_\mathrm{k}>5$~TeV/n in the two above cases,
respectively. A detection of this flat energy-dependent ratio would be
consistent with GCR production in SNRs.

The actual mean galactic escape length $x$ is smaller than in the models
which neglect the production secondary nuclei inside the sources and only
take into account spallation during the random walk of primaries in
the Galactic disk after release from their sources. The extent of this
reduction depends upon the mean ISM density at the SNR sites: $x$ is from
10\% to 30\% smaller if SNRs on average explode in an ISM with hydrogen
number density $N_\mathrm{H}$ between 0.3 and 1~cm$^{-3}$, respectively. This
effect is much smaller compared with the estimates of Wandel (\cite{wan88}), who
suggested efficient CR acceleration/reacceleration by much older and
weaker SNR shocks. Only if the ISM number density is as low as
$N_\mathrm{H}=0.003$~cm$^{-3}$ one expects the mean escape length to be smaller by
a factor of 3 compared with the standard models.

\begin{acknowledgements} 
The authors are grateful to the International Space Science Institute
(Bern) for its financial support of their work in the team "Energetic
particles in the Galaxy: acceleration, transport and gamma-ray
production". EGB and LTK acknowledge the hospitality of the
Max-Planck-Institut f\"ur Kernphysik, where part of this work was
carried out.  The work has been supported in part by the Russian
Foundation for Basic Research (grants 03-02-16325;  01-02-17460).
\end{acknowledgements}


\begin{thebibliography}{99}
 
\bibitem[1982]{abbott}
Abbott, D.C. 1982, Ape, 263, 723

\bibitem[1996]{ber96}
Berezhko, E.G. 1996, Astropart.\ Phys., 5, 367

\bibitem[1997]{bv97}
Berezhko, E.G. and V\"olk, H.J. 1997, Astropart. Phys., 7, 183

\bibitem[2000]{bv00}
Berezhko, E.G. and V\"olk, H.J. 2000, A\&A, 357, 183

\bibitem[1996]{byk96}
Berezhko, E.G., Yelshin, V.K., Ksenofontov, L.T. 1996, JETP 82, 1

\bibitem[1999]{bk99}
Berezhko, E.G., Ksenofontov, L.T. 1999, JETP 116, 737
 
\bibitem[1990]{brz90}
Berezinskii, V.S., Bulanov, S.V., Dogiel, V.A., Ginzburg, V.L., Ptuskin, V.S., 
1990, Astrophysics of Cosmic Rays (Amsterdam: North Holland)
 
\bibitem[1980]{bo80}
Blandford, R.D., Ostriker, J.P. 1980, ApJ 237, 793.

\bibitem[1996]{dv96}
Dorfi, E.A. \& V\"olk, H.J. 1996, A\&A, 307, 715 

\bibitem[1990]{eng90}
Engelmann, J.J. et al. 1990, A\&A, 233, 96.

\bibitem[1969]{haya69}
Hayakawa, S., 1969, Cosmic-Ray Physics (New York: Wiley Interscience)
 
\bibitem[2001]{jlpw01}
Jones, F.C., Lukasiak, A., Ptuskin, V.S., Webber, W.R., 2001, ApJ 547, 264 
 
\bibitem[1987]{wel87}
Wandel, A., Eichler, D., Letaw, J.R. et al., 1987, ApJ 316, 676.

\bibitem[1988]{wan88}
Wandel, A. 1988, A\&A, 200, 279.

\bibitem[2003]{vbk03}
V\"olk, H.J., Berezhko, E.G. \& Ksenofontov, L.T. 2003 {\it to appear in 
in A\&A}, astro-ph/0306016.

\end{thebibliography}
\end{document}